\documentclass{iaus}
\usepackage{graphicx}

\def\LCDM{{$\Lambda$CDM}}
\def\spose#1{\hbox to 0pt{#1\hss}}
\def\la{\mathrel{\spose{\lower.5ex\hbox{$\mathchar"218$}}
     \raise.4ex\hbox{$\mathchar"13C$}}}
\def\ga{\mathrel{\spose{\lower.5ex \hbox{$\mathchar"218$}}
     \raise.4ex\hbox{$\mathchar"13E$}}}

\def\aap{{\it A}\&{\it A}}
\def\aj{{\it AJ}}
\def\apj{{\it Ap.\ J.}}
\def\apjl{{\it Ap.\ J. Lett.}}
\def\araa{{\it Ann.\ Rev.\ Astron.\ Ap.}}
\def\lrr{{Liv.\ Rev.\ Rel.}}
\def\mnras{{\it MNRAS}}
\def\nat{{\it Nature}}
\def\prl{{\it Phys.\ Rev.\ Lett.}}
\def\sci{{\it Science}}

\title[IAU Symposium 254.~~Dark Matter Density] 
{Dark Matter Density in Disk Galaxies}

\author[Sellwood]{J. A. Sellwood}
\affiliation{Dept. of Physics \& Astronomy, Rutgers University \\
136 Frelinghuysen Road, Piscataway, NJ 08855 \\
email: {\tt sellwood@physics.rutgers.edu}}

\pubyear{2008}
\volume{254}  
\pagerange{1--12}
\jname{The Galaxy Disk in Cosmological Context}
\editors{J. Andersen, J. Bland-Hawthorn \& B. Nordstr\"om}
\begin{document}

\maketitle

\begin{abstract}
I show that the predicted densities of the inner dark matter halos in
\LCDM\ models of structure formation appear to be higher than
estimates from real galaxies and constraints from dynamical friction
on bars.  This inconsistency would not be a problem for the \LCDM\
model if physical processes that are omitted in the collisionless
collapse simulations were able to reduce the dark matter density in
the inner halos.  I review the mechanisms proposed to achieve the
needed density reduction.

\end{abstract}

\firstsection 
\section{Motivation}
I was invited to review secular evolution in disk galaxies.  Rather
than attempt a very superficial review of this vast topic, I here
focus on dynamical friction.  Several other possible topics could be
included in a review of secular evolution, such as: scattering of disk
stars, which I reviewed only recently (Sellwood 2008a); mixing and
spreading of disks (\eg~Sellwood \& Binney 2002; Ro\u skar \etal\ 2008;
Freeman, these proceedings); and the formation of pseudo-bulges (\eg\
Kormendy \& Kennicutt 2004; Binney, these proceedings).

The current \LCDM\ paradigm for galaxy formation (\eg~White, these
proceedings) makes specific predictions for the dark matter (DM)
densities in halos of galaxies.  I first argue that halos of some
barred galaxies are inconsistent with this prediction, and then
consider whether DM halo densities could be lowered by internal galaxy
evolution.

\section{Inner Halo Density}
Attempts to measure the halo density and its slope in the innermost
parts of galaxies are beset by many observational and modeling issues
(\eg~Rhee \etal\ 2004; Valenzuela \etal\ 2007), while the predictions
from simulations in the same innermost region are still being revised,
as shown earlier by White.  It therefore makes sense to adopt a more
robust measure of central density, such as that proposed by Alam,
Bullock \& Weinberg (2002).  Their parameter, $\Delta_{v/2}$, is a
measure of the mean DM density, normalized by the cosmic closure
density, interior to the radius at which the circular rotational speed
due the DM alone rises to half its maximum value.  For those more
familiar with halo concentrations, it is useful to note that for the
precise NFW (Navarro, Frenk \& White 1997) halo form, $\Delta_{v/2} =
672 c^3 /[\ln(1+c) - c/(1+c)]$, if $c$ is defined where the mean halo
density is 200 times the cosmic closure value; thus $\Delta_{v/2}
\simeq 10^{5.5}$ for a $c=9$ halo.  However, a further advantage of
$\Delta_{v/2}$ is that it is not tied to a specific density profile.

\begin{figure}[t]
\begin{center}
\includegraphics[width=10cm]{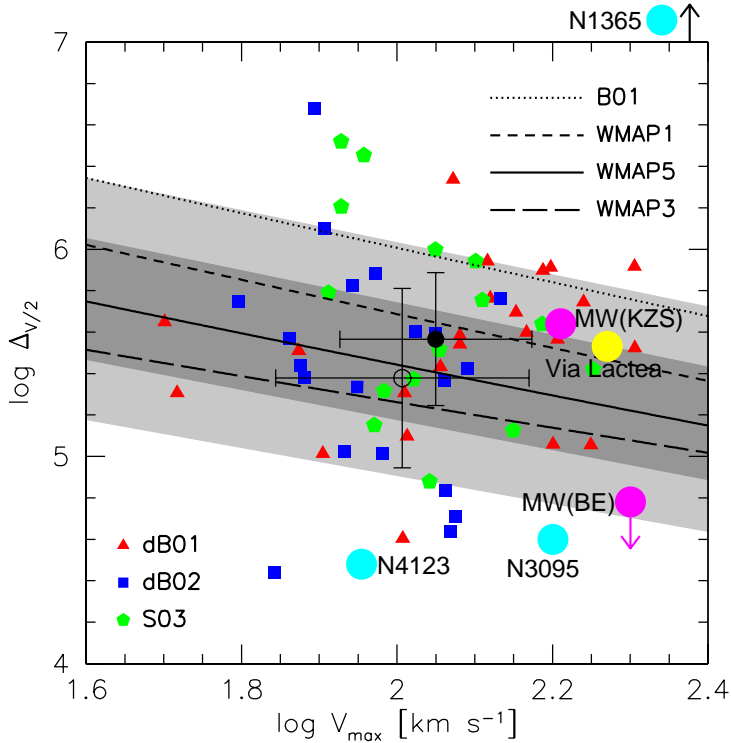} 
\caption{Figure reproduced from Macci\'o \etal\ (2008, with
permission), to which I have added the large labeled points that are
described in the text.  The shaded regions show the $1-$ and
$2-\sigma$ ranges of the predicted values of $\Delta_{v/2}$, while the
lines show the means, as functions of the maximum circular speed from
the DM halos.  The small colored symbols show various estimates of
these parameters for dwarf and LSB galaxies estimated by Macci\'o
\etal\ from data in de Blok, McGaugh \& Rubin (2001), de Blok \& Bosma
(2002), and Swaters \etal\ (2003).}
\label{Delfig}
\end{center}
\end{figure}

\subsection{Prediction}
Figure ~\ref{Delfig}, reproduced from Macci\'o, Dutton \& van den
Bosch (2008), shows the \LCDM\ prediction (shaded) that results when
the initial amplitude and spectrum of density fluctuations match the
latest cosmic parameters, as determined by the WMAP team (Komatsu
\etal\ 2008).  I have added one further predicted point from the {\it
Via Lactea\/} model (Diemand, Kuhlen \& Madau 2007), which is argued
to resemble a typical halo that would host a galaxy such as the Milky
Way.

\subsection{Data from Galaxies}
Macci\'o \etal\ plot the small colored triangles, squares and
pentagons, which show estimates of $\Delta_{v/2}$ culled from the
literature.  The data are from dwarf and LSB galaxies, that are
believed to be DM dominated and the large black circle indicates
their mean in both coordinates, with the error bars indicating
the ranges.  The open circle with error bars shows a revised mean
after subtracting a contribution to the central attraction by the
estimated baryonic mass in these galaxies.  These authors conclude
that these data are consistent with the model predictions.

The large cyan circles are estimates for large galaxies: both NGC~4123
(Weiner, Sellwood \& Williams 2001) and NGC~3095 (Weiner 2004) have
well-estimated halos that are significantly below the predicted range.
The inner density estimated for NGC~1365 (Zanm\'ar Sanch\'ez \etal\
2008), on the other hand, is $\Delta_{v/2} \sim 5 \times 10^7$, which
is off the top of this plot, though the total halo mass is quite
modest; a large uncertainty in the inclination, a possible warp that
is very hard to model, together with evidence of a inner disturbance
that required us to fit to data only one side of the bar, all conspire
to render the inner halo density of this Fornax cluster galaxy quite
uncertain.

I also plot two magenta points from different mass models for the
Milky Way.  The upper point is from Klypin, Zhao \& Somerville (2002)
while the lower shows the upper bound on the inner halo density
estimated by Binney \& Evans (2001).  Their bound comes from trying to
include enough foreground disk stars to match an old estimate
(Popowski \etal\ 2001) of the micro-lensing optical depth to the
red-clump stars of the Milky Way bulge; current estimates of this
optical depth are somewhat lower (Popowski \etal\ 2005), suggesting a
reanalysis will allow a higher halo density.

It is important to realize that the creation of a disk through
condensation, or inflow, of gas into the centers dark halos must
deepen the gravitational potential and cause the halo to contract
(\eg~Blumenthal \etal\ 1986; Sellwood \& McGaugh 2005).  Thus,
estimates of the current halo density should be reduced to take
account of halo compression.  Allowance for compression brings the
point for NGC~1365 down by more than one order of magnitude.  But
making this correction for all the galaxies (which Macci\'o \etal\ did
not do for their open circle point) will move all the data points
down, including the well-estimated points for NGC~4123 \& NGC~3095
that are already uncomfortably low.

The only real difficulties presented by the comparison with the
predictions in Fig.~1 arise from two well-determined low points, which
could simply turn out to be anomalous.  Additional evidence suggesting
uncomfortably low DM densities in real galaxies comes from other
rotation curve data (\eg\ Kassin, de Jong \& Weiner 2006) and the
difficulty of matching the observed zero point of the Tully-Fisher
relation (\eg\ Dutton, van den Bosch \& Courteau 2008).  However, an
independent argument, based on the constraints from dynamical friction
on bars, also suggests that the DM density in barred galaxies is
generally lower than predicted.

\subsection{Bar Slow Down}
\label{slow}
Bars in real galaxies are generally believed to be ``fast'', in that
the radius of corotation is generally larger than the semi-major axis
of the bar by only a small factor, $\cal R$. Indications that $1 \la
{\cal R} \la 1.3$ come from (a) direct measurements in largely
gas-free galaxies, summarized by Corsini (2008), (b) models of the gas
flow (\eg~Weiner \etal\ 2001; Bissantz, Englmaier \& Gerhard 2003),
and (c) indirect arguments about the location of dust lanes
(\eg~Athanassoula 1992).  Rautiainen, Salo \& Laurikainen (2008), and
others, claim a few counter-examples from indirect evidence, although
they concede that they try to match the morphology of the spiral
patterns, which may rotate more slowly than the bar.

After some considerable debate, a consensus seems to be emerging that
strong bars in galaxies should experience fierce braking unless the
halo density is low (Debattista \& Sellwood 1998, 2000; O'Neill \&
Dubinski 2003; Holley-Bockelmann, Weinberg \& Katz 2005; Col\'\i n,
Valenzeula \& Klypin 2006).  The counter-example claimed by Valenzuela
\& Klypin (2003) was shown by Sellwood \& Debattista (2006) to have
resulted from a numerical artifact in their code.  The claims of
discrepancies by Athanassoula (2003) are merely that weak, or
initially slow bars, are less strongly braked, while she also finds
that strong, fast bars slow unacceptably in dense halos.

\begin{figure}[t]
\begin{center}
\includegraphics[width=10cm]{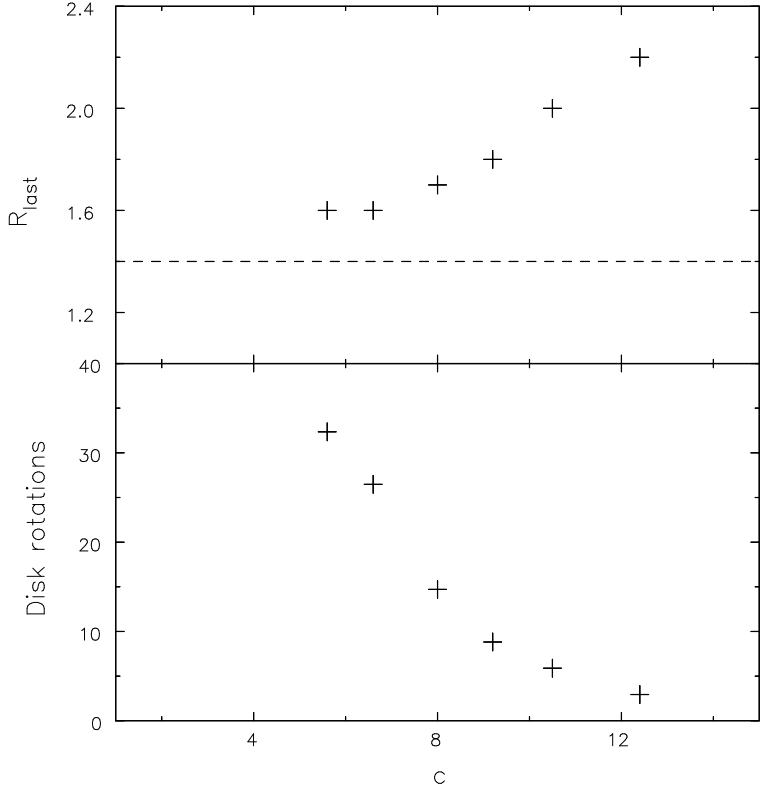} 
\end{center}
\caption{Bars in NFW halos.  Above: The value of $\cal R$ at the
time each simulation was stopped.  Below: The number of disk rotations
before the bar slowed to the point where ${\cal R}=1.4$.}
\label{rplot}
\end{figure}

Thus there is little escape from the conclusion by Debattista \&
Sellwood (2000), that the existence of fast bars in strongly barred
galaxies requires a low density of DM in the inner halo.  Our original
constraint required near maximal disks, although the halo models in
that paper were not at all realistic.  Figure~\ref{rplot} summarizes
the results from a new study of exponential disks embedded in NFW
halos, computed using the code described in Sellwood (2003) that has
greatly superior dynamic range.  When scaled to the Milky Way, a
rotation period at 3 disk scale lengths in these models is 270~Myr.
In all cases, the bar becomes slow by the end of the experiment,
although the number of disk rotations needed until ${\cal R} > 1.4$
increases as the concentration index, $c$, is reduced.  Thus, friction
in NFW halos causes bars to become unacceptably slow in a few disk
rotations when $c \ga 10$, but on a time scale $\ga 7.5\;$Gyr when $c
\la 6$ or when the uncompressed $\Delta_{v/2} \la 10^{5.1}$.  This
conservative bound would exclude well over half the predicted range of
halo densities in Fig.~\ref{Delfig} for an uncompressed $V_{\rm max} =
10^{2.25} \simeq 180\;$km/s.

In the context of this symposium, it would be nice to test Milky Way
models for bar slow-down.  The halo and disk in the model tested by
Valenzuela \& Klypin (2003) were selected from the Klypin \etal\
(2002) models for the Milky Way.  The simulations reveal that a very
large bar with semi-major axis $\ga 5\;$kpc forms quickly, which slows
unacceptably within $\sim 5\;$Gyr.  Unfortunately, the absence of a
realistic bulge in these experiments crucially prevents these results
from being regarded as a test of the Klypin \etal\ (2002) MW models,
since a bulge should cause a much shorter and faster bar to form.

\section{Can the DM Density Be Reduced?}
The \LCDM\ model would not be challenged if the present-day DM density
in galaxies can be reduced by processes that are neglected in cosmic
structure formation simulations, which generally follow the dynamics
of collisionless collapse only.  Four main ideas have been advanced
that might achieve the desired density decrease.

\subsection{Feedback}
This first idea is not a secular effect, and therefore strictly falls
outside my assigned topic.  However, I discuss it briefly because it
should not be omitted from any list of processes that might effect a
density reduction.

The basic idea, proposed by Navarro, Eke \& Frenk (1997), Binney,
Gerhard \& Silk (2001), and others is that gas should first collect
slowly in a disk at the center of the halo, thereby deepening the
gravitational potential well and compressing the halo adiabatically.
A burst of star formation would then release so much energy that most
of the gas would be blasted back out of the galaxy at very high speed,
resulting in a non-adiabatic decompression of the halo, which may
possibly result in a net reduction in DM density.

Gnedin \& Zhao (2002) present the definitive test of the idea.  In
their simulations, they slowly grew a disk inside the halo, causing it
to compress adiabatically, and then they instantaneously removed the
disk.  With this artifice, they deliberately set aside all questions
of precisely how the star burst could achieve the required outflow, in
order simply to test the extreme maximum that any conceivable feedback
process could achieve.

They found that the final density of the halo was lower than the
initial, confirming that the effect can work, and that mass blasted
out from the very center of the potential has greatest effect,
presumably because it produces the largest instantaneous change in the
gravitational potential.  However, density reductions by more than a
factor of two required that the disk be unreasonably concentrated, and
consequently the baryonic mass has to be blasted out from deep in the
potential well.

\subsection{Bar-Halo Friction}
The same physical process that slows bars, discussed in \S\ref{slow},
can also reduce the density of the material that takes up the angular
momentum, as first reported by Hernquist \& Weinberg (1992).  This
mechanism prompted Weinberg \& Katz (2002) to propose the following
sequence of events as a means to reconcile \LCDM\ halo predictions
with bar pattern speed constraints and other data.  They argued that a
large bar in the gas at an early stage of galaxy formation could
reduce the DM density through dynamical friction.  The gas bar would
then disperse as star formation proceeded, so that were a smaller
stellar bar to form later it would not experience much friction.
Their idea has been subjected to intensive scrutiny.

Normal Chandrasekhar friction (\eg~Binney \& Tremaine 2008, \S8.1) is
formally invalid in more realistic dynamical systems, such as
quasi-spherical halos, because the background particles are bound to
the system and will return to interact with the perturber repeatedly.
Tremaine \& Weinberg (1984) showed that under these circumstances
angular momentum exchange occurs at resonances between the motion of
the perturber and that of the background particles.

The $N$-body simulations mentioned in \S\ref{slow} generally did not
produce a substantial reduction in halo density, despite the presence
of strong friction.  This could be because the bars were not strong
enough, but Weinberg \& Katz (2007a,b) argue that the simulations were
too crude and that delicate resonances would not be properly mimicked
in simulations unless the number of particles exceeds between $10^7$
\& $10^9$, depending on the bar size and strength and the halo mass
profile.

Thus two major questions arise: (1) are results from simulations
believable? and (2) can realistic bars cause a large density decrease?
I addressed both these issues in a recent paper (Sellwood 2008b).
While rigid, ellipsoidal bars are not terribly realistic, I used them
deliberately in order to test the analysis and to compare with the
simulations presented by Weinberg \& Katz.  Dubinski (these
proceedings) presents results of similar tests using fully
self-consistent disks that form bars.

\subsubsection{What $N$ Is Enough?}
Simple convergence tests reveal that experiments with different
numbers of particles converge to an invariant time evolution of both
the pattern speed and halo density changes at quite modest numbers of
halo particles.  I report that $N=10^5$ seemed to be sufficient for a
very large bar, while $N\sim 10^6$ was needed for a more realistic
bar.  I observed no change the results in either case as I increased
to $N=10^8$, or when I employed a spectrum of particle masses in order
to concentrate more into the crucial inner halo.  I found results for
different numbers of particles overlaid each other perfectly, with no
evidence for the stochasticity that Weinberg \& Katz predicted should
result if few particles were in resonance.

I also demonstrated that my simulations did indeed capture resonant
responses that converged for the same modest particle numbers.  I
measured the change in the density of particles $F(L_{\rm res})$,
where $L_{\rm res}$ is an angular momentum-like variable that depends
on orbit precession frequency.  Using this variable, I was able to
estimate that some 7\% -- 20\% of halo particles participated in
resonant angular momentum exchanges with the bar during a short time
interval.  This fraction is vastly greater than Weinberg \& Katz
predicted, because they neglected to take into account the broadening
of resonances caused by the evolving bar perturbation, that both grows
and slows on an orbital timescale.  Athanassoula (2002) and Ceverino
\& Klypin (2007) also demonstrated the existence of resonances in
their simulations.

\begin{figure}[t]
\begin{center}
\includegraphics[width=12cm]{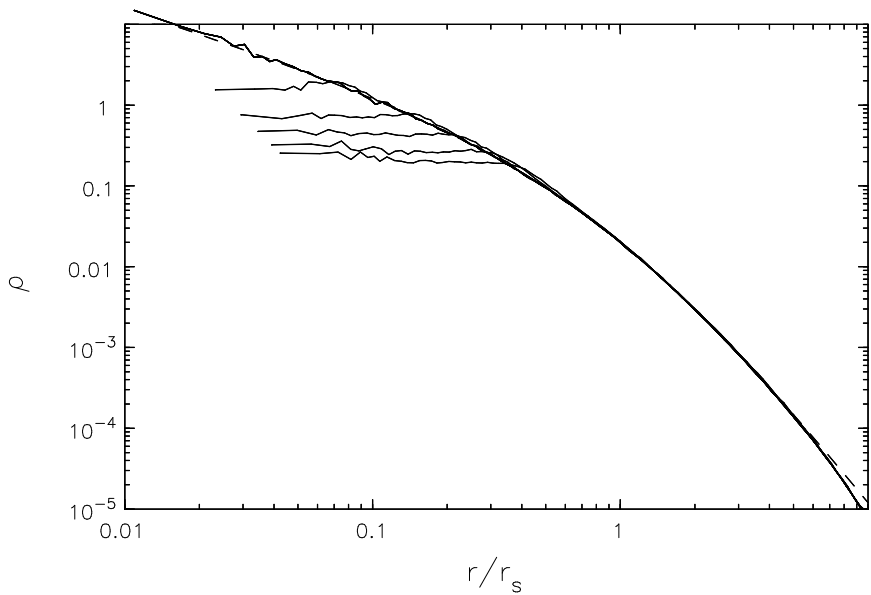} 
\caption{Results from five different experiments with
different bar lengths.  The dashed line shows the initial profile,
while the solid lines show estimates from the particles of the initial
(cusped) and final (cored) density profiles from a series of runs with
different bar semi-major axes.  The final density lines from the lowest
to the highest are for bar lengths, $a/r_s = 1$, 0.8, 0.6, 0.4, \& 0.2.}
\label{length}
\end{center}
\end{figure}

\subsubsection{Density Reduction by Very Strong Bars}
Figure~\ref{length} shows that a strong bar rotating in a halo within
a density cusp ($\rho \propto 1/r$) can flatten the cusp to $\sim 1/3$
bar length.  The rigid bar needed to accomplish this must have an axis
ratio $a/b \ga 3$, a mass $M_b \ga 30\%$ of the halo mass inside $r = a$.

While cusp flattening is a driven response caused by the slowing bar,
it is also a collective effect.  I find a much smaller change when I
hold fixed the monopole terms of the halo self-gravity.  Thus it is
dangerous, when studying halo density changes, to include any rigid
mass component.

The only simulation I am aware of in which a self-consistent bar
flattened the inner density cusp is that reported by Holley-Bockelmann
\etal\ (2005).  They report a significant density reduction that
flattened the cusp to a radius $\sim a/5$.  Note that in their model,
the initial halo density was not compressed by the inclusion of the disk,
since they rederived the halo distribution function that would be in
equilibrium in the potential of an uncompressed NFW halo plus the disk.

\begin{figure}[t]
\begin{center}
\includegraphics[width=10cm]{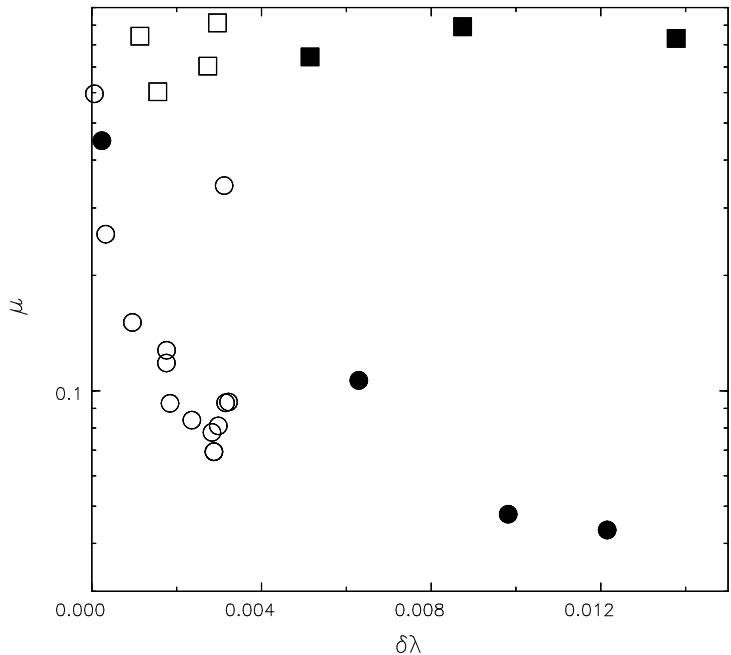}
\caption{Fractional changes, $\mu$, to $\Delta_{v/2}$ in many
experiments. The abscissae show the angular momentum given to the
halo, expressed as the usual dimensionless spin parameter.  Open
circles mark results from experiments in which the density profile of
the inner cusp was flattened, while squares indicate experiments where
cusp flattening did not occur.  Filled symbols show results from
experiments in which the moment of inertia of the bar was increased by
a factor 5 in all cases except the point at the upper right, where the
MoI was increased 10-fold (for more details see Sellwood 2008).  The
changes to $\Delta_{v/2}$ make no allowance for halo compression.}
\label{ABW}
\end{center}
\end{figure}

\subsubsection{More Gradual Changes}
A number of other simulations in the literature have revealed a modest
density reduction caused by angular momentum exchange with a bar in
the disk.  \eg~Debattista \& Sellwood (2000) show a reduction in the
halo contribution to the central attraction, and something similar can
also be seen in Athanassoula's (2003) simulations.  None of these
models included very extensive halos.

The inner halo density in fully self-consistent simulations with more
extensive and cusped halos can actually rise as the model evolves
(Sellwood 2003; Col\'\i n \etal\ 2006).  This happens because angular
momentum lost by the bar in the disk causes it to contract; the
deepening potential of the disk causes further halo compression that
overwhelms any density reduction resulting from the angular momentum
transferred to the halo.

\subsubsection{Angular Momentum Reservoir}
A crucial consideration that limits the magnitude of halo density
reduction by bar friction is the total angular momentum available in
the baryonic disk.  Tidal torques in the early universe lead to halos
with a log-normal distribution of spin parameters with a mean $\lambda
\sim 0.05$, where the dimensionless spin parameter is $\lambda =
LE^{1/2}/GM^{5/2}$ as usual.  Assuming that the baryons and dark
matter are well mixed initially, the fraction of angular momentum in
the baryons is equal to the baryonic mass fraction in the galaxy: some
5\% -- 15\%.  Thus total angular momentum loss from the disk could
increase the halo spin parameter by typically $\delta\lambda \sim
0.005$.

Figure~\ref{ABW}, taken from Sellwood (2008b), shows the factor by
which the halo density is reduced, $\mu = \Delta_{v/2,\rm fin} /
\Delta_{v/2,\rm init}$ as the ordinate against the angular momentum
gain of halo.  A density reduction by a factor of 10 is possible, but
the bar must be extreme, having a semi-major axis, $a$, approaching
that of the break radius, $r_s$, of the NFW profile and a mass $>30\%$
halo interior to $r = a$.  Furthermore, such a density reduction is
achieved at the expense of removing a large fraction of the angular
momentum of the baryons.  It should also be noted that the density
changes shown in Fig.~\ref{ABW} also do not take account of any halo
compression that might have occurred as the bar and disk formed.

\subsection{Baryonic Clumps}
El-Zant, Shlosman \& Hoffman (2001) proposed that dynamical friction
from the halo on moving clumps of dense gas will also transfer energy
to the DM and lower its density.  They envisaged that baryons would
collect into clumps through the Jeans instability as galaxies are
assembled and present somewhat simplified calculations of the
consequences of energy loss to the halo.  The idea was taken up by Mo
\& Mao (2004), who saw this as a means to erase the cusps in small
halos before they merge to make a main galaxy halo, and by Tonini,
Lapi \& Salucci (2006).

The proposed mechanism has a number of conceptual problems, however.
The model assumes that the settling gas clumps maintain their
coherence for many dynamical crossing times without colliding with
other clumps or being disrupted by star formation, for example.  In
addition, calculations (\eg~Kaufmann \etal\ 2006) of the masses of
condensing gas clumps suggest they range up to only $\sim
10^6\;M_\odot$, which is too small to experience strong friction.
Larger clumps will probably gather in subhalos, which may get dragged
in, but simulations with sub-clumps composed of particles (\eg~Ma \&
Boylan-Kolchin 2004) indicate that the DM halos of the clumps will be
stripped, which simply replaces any DM moved outwards in the halo.
Debattista \etal\ (2008) suggest halo compression is an issue here
also, but the essential idea suggested by El-Zant \etal\ is to
displace the DM as the gas settles, which avoids halo compression.

\begin{figure}[t]
\begin{center}
\includegraphics[width=12cm]{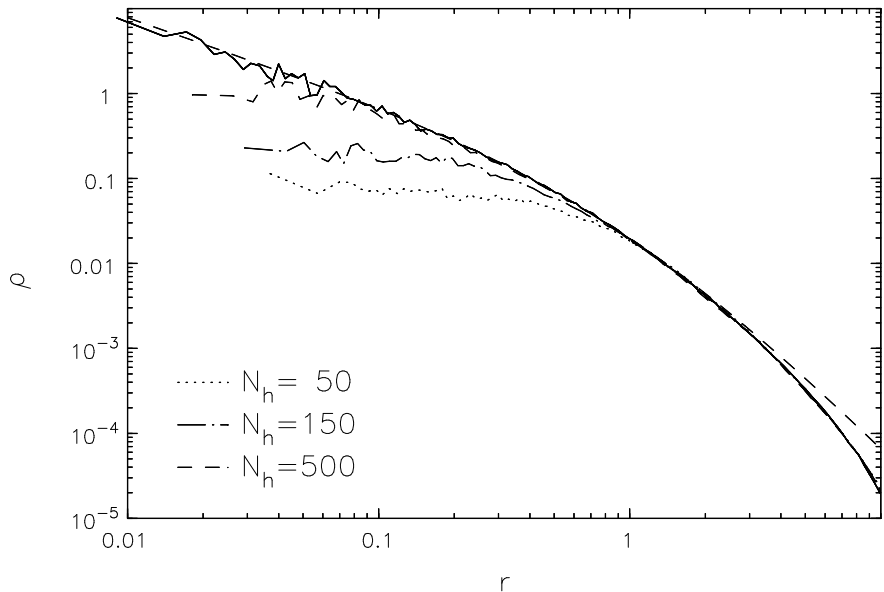}
\caption{The changes in density caused by the settling of $N_h$ heavy
particles with total mass of $0.1M_{200}$, initially distributed at
uniform density within a sphere of radius $4rs$ in an NFW halo.  The
solid line shows the density measured from the particles at the start
while the broken lines show the density after 9 Gyr.  Another dashed
line shows the corresponding theoretical NFW curve.}
\label{collect}
\end{center}
\end{figure}

\subsubsection{A Direct Test}
Setting all these difficulties to one side, Jardel \& Sellwood (2008)
set out to test the mechanism with $N$-body simulations.  As proposed
by El-Zant \etal, we divided the entire mass of baryons into $N_h$
equal mass clumps, treated as softened point masses, to which we added
isotropic random motion to make their distribution in rough dynamical
equilibrium inside an NFW halo composed of $\sim 1\,$M
self-gravitating particles.  All particles, both light and heavy,
experienced the attraction of all others.

Figure~\ref{collect} shows results after $\sim 9\;$Gyr, when scaled to
a $c=15$ halo -- the timescale would be even longer for less
concentrated halos.  We find that some density reduction does occur,
but the rate at which the density is decreased is considerably slower
than El-Zant \etal\ predict.  We traced this discrepancy to their use
of three times too large a Coulomb logarithm in their calculations.

Fig.~\ref{collect} also shows that the rate of density reduction rises
as the baryon mass is concentrated into fewer, more massive particles.
Again our result is consistent with that of Ma \& Boylan-Kolchin
(2004), who employed a mass spectrum of clumps, and who showed that a
much smaller density reduction occurred in a separate simulation that
omitted the three heaviest clumps.  Thus, if this process is to work
on an interesting time-scale, it requires a few gas clumps whose
masses exceed $1\%$ of the entire halo.

Mashchenko \etal\ (2006, 2007) argue that the energy input to the
halo, mediated by the motion of the mass clumps, can be boosted if the
gas is stirred by stellar winds and supernovae -- a less extreme form
of feedback ({\it cf.}~\S3.1).  Their simulations of the effect reveal
a density reduction in dwarf galaxies.  Peirani \etal\ (2008), on the
other hand, propose AGN activity to accelerate gas clumps.  They
present simulations that show the cusp can be flattened to $\sim
0.1r_s$ with a clump having mass of $\sim1$\% of the galaxy mass,
being driven outwards from the center to a distance of half the NFW
break radius at a speed of 260 km/s.  In both these models, it is
unclear how the dense material can be accelerated to the required
speed (\eg~MacLow \& Ferrara 1999).

\subsection{Recoiling/Binary BHs}
In any hierarchical structure-formation model, halos grow through a
succession of mergers (\eg~Wechsler \etal\ 2002).  If massive black
holes (BHs) have formed in the centers of two galaxies that merge,
then one expects both BHs to settle to the center of the merged halo
and to form a binary pair of BHs in orbit about each other.  The
physics of the decay of the orbit is interestingly complicated
(\eg~Merritt \& Milosavljevi\'c 2005).

The star density in the centers of elliptical galaxies can be reduced
by star scattering as the BH binary hardens, and also by the separate
process of BH recoil if the binary encounters another massive object.
Merritt \& Milosavljevi\'c (2005) point out that the star density can
be significantly reduced only within the sphere of gravitational
influence of the BHs, which extends to $r \sim r_h$, where $r_h =
GM_\bullet/\sigma^2$, where $M_\bullet$ is the mass of the BH and
$\sigma$ is the velocity dispersion of the stars.  Generous values
for disk galaxies might be $M_\bullet = 10^7 M_\odot$ and $\sigma =
100\;$km/s, yielding $r_h \simeq 4.4\;$pc.

Since none of the processes that affect the star density in the center
depend upon the masses of the individual stars, DM particles will be
affected in a similar manner, and we must expect the DM density to be
depleted also over the same volume.  However, because black hole
dynamics affects only the inner few parsecs, it can have essentially
no effect on bar pattern speed constraints or values of parameters
such as $\Delta_{v/2}$.

\section{Conclusions}
The inner densities of DM halos of galaxies today continue to
challenge the \LCDM\ model for galaxy formation.  Both direct
estimates in a few galaxies, and dynamical friction constraints from
bar pattern speeds require inner densities lower than predicted in
DM-only simulations by at least a factor of a few.

Several processes that are omitted from DM only simulations can reduce
the inner halo density.  Feedback has a very slight effect unless a
large mass of gas is blasted out from the deepest point in the
peotential well.  Bar friction requires an extreme bar and removes a
large fraction of the angular momentum in the baryons.  Dynamical
friction on massive gas clumps is too slow, unless moving gas clumps
exceed $\sim 1\%$ of total baryonic mass.  Scattering by merging or
recoiling BHs affects only very center.  Thus any of the four
suggested mechanism needs to be stretched if it is to cause a
significant reduction.

\acknowledgments I thank Victor Debattista for comments on the
manuscript and the organizers of the conference for travel support.
This work was supported by grants AST-0507323 from the NSF and
NNG05GC29G from NASA.

\section*{References}

\smallskip
\parindent=0pt
\everypar{\hangindent 1.2cm}
\footnotesize

Alam, S. M. K., Bullock, J. S. \& Weinberg, D. H. 2002, \apj, {\bf 572}, 34

Athanassoula, E. 1992, \mnras, {\bf 259}, 345

Athanassoula, E. 2002, \apjl, {\bf 569}, L83

Athanassoula, E. 2003, \mnras, {\bf 341}, 1179

Binney, J. J. \& Evans, N. W. 2001, \mnras, {\bf 327}, L27

Binney, J., Gerhard, O. \& Silk, J. 2001, \mnras, {\bf 321}, 471

Binney, J. \& Tremaine, S. 2008, {\it Galactic Dynamics\/} 2nd Ed.\ (Princeton: Princeton University Press)

Bissantz, N., Englmaier, P. \& Gerhard, O. 2003, \mnras, {\bf 340}, 949

Blumenthal, G. R., Faber, S. M., Flores, R. \& Primack, J. R. 1986, \apj, {\bf 301}, 27

Ceverino, D. \& Klypin, A. 2007, \mnras, {\bf 379}, 1155

Col\'\i n, P., Valenzuela, O. \& Klypin, A. 2006, \apj, {\bf 644}, 687

Corsini, E. M. 2008, in {\it Formation and Evolution of Galaxy Disks}, eds.\ J. G. Funes SJ \& E. M. Corsini (ASP, to appear)

Debattista, V. P. \& Sellwood, J. A. 1998, \apjl, {\bf 493}, L5

Debattista, V. P. \& Sellwood, J. A. 2000, \apj, {\bf 543}, 704

Debattista, V. P., \etal\ 2008, \apj, {\bf 681}, 1076

de Blok, W. J. G., McGaugh, S. S. \& Rubin, V. C. 2001, \aj, {\bf 122}, 2396

de Blok, W. J. G. \& Bosma, A. 2002, \aap, {\bf 385}, 816

Diemand, J., Kuhlen, K. \& Madau, P. 2007, \apj, {\bf 667}, 859

Dutton, A. A., van den Bosch, F. C. \& Courteau, S. 2008, in {\it Formation and Evolution of Galaxy Disks}, eds.\ J. G. Funes SJ \&  E. M. Corsini (ASP, to appear) arXiv:0801.1505

El-Zant, A., Shlosman, I. \& Hoffman, Y. 2001, \apj, {\bf 560}, 636

Gnedin, O. Y. \& Zhao, H.S., 2002, \mnras, {\bf 333}, 299

Hernquist, L. \& Weinberg, M. D. 1992, \apj, {\bf 400}, 80

Holley-Bockelmann, K., Weinberg, M. \& Katz, N. 2005, \mnras, {\bf 363}, 991

Jardel, J. \& Sellwood, J. A. 2008, \apj, (submitted)

Kassin, S. A., de Jong, R. S. \& Weiner, B. J. 2006, \apj, {\bf 643}, 804

Kaufmann, T., Mayer, L., Wadsley, J., Stadel, J. \& Moore, B. 2006, \mnras, {\bf 370}, 1612

Klypin, A., Zhao, HS. \&  Somerville, R. S. 2002, \apj, {\bf 573}, 597

Komatsu, E. \etal\ 2008, arXiv:0803.0547

Kormendy, J. \& Kennicutt, R. C. 2004, \araa, {\bf 42}, 603

Ma, C-P. \& Boylan-Kolchin, M. 2004, \prl, {\bf 93}, 21301

Macci\`o, A. V., Dutton, A. A. \& van den Bocsh, F. C. 2008, arXiv:0805.1926

MacLow, M-M. \& Ferrara, A. 1999, \apj, {\bf 513}, 142

Mashchenko, S., Couchman, H. M. P. \& Wadsley, J. 2006, \nat, {\bf 442}, 539

Mashchenko, S., Couchman, H. M. P. \& Wadsley, J. 2007, \sci, {\bf 319}, 174

Merritt, D. \& Milosavljevi\'c, M. 2005, \lrr, {\bf 8}, 8 (astro-ph/0410364)

Mo, J. J. \& Mao, S. 2004, \mnras, {\bf 353}, 829

Navarro, J. F., Eke, V. R. \& Frenk, C. S. 1997, \mnras, {\bf 283}, L72

Navarro, J. F., Frenk, C. S. \& White, S. D. M. 1997, \apj, {\bf 490}, 493

O'Neill, J. K. \& Dubinski, J. 2003, \mnras, {\bf 346}, 251

Peirani, S, Kay, S. \& Silk, J. 2008, \aap, {\bf 479}, 123

Popowski, P. \etal\ 2001, in {\it Astrophysical Ages and Times Scales}, eds.\ T. von Hippel, C. Simpson, \& N. Manset, ASP Conference Series {\bf 245}, p.~358

Popowski, P. \etal\ 2005, \apj, {\bf 631}, 879

Rautiainen, P., Salo, H. \& Laurikainen, E. 2008, arXix:0806.0471

Rhee, G., Valenzuela, O., Klypin, A., Holtzman, J. \& Moorthy, B. 2004, \apj, {\bf 617}, 1059

Ro\u skar, R., Debattista, V. P., Stinson, G. S., Quinn, T. R., Kaufmann, T. \& Wadsley, J. 2008, \apjl, {\bf 675}, L65

Sellwood, J. A. 2003, \apj, {\bf 587}, 638

Sellwood, J. A. 2008a, in {\it Formation and Evolution of Galaxy Disks}, eds.\ J. G. Funes SJ \&  E. M. Corsini (ASP, to appear) arXiv:0803.1574

Sellwood, J. A. 2008b, \apj, {\bf 679}, 379

Sellwood, J. A. \& Binney, J. J. 2002, \mnras, {\bf 336}, 785

Sellwood, J. A. \& Debattista, V. P. 2006, \apj, {\bf 639}, 868

Sellwood, J. A. \& McGaugh, S. S. 2005, \apj, {\bf 634}, 70

Swaters, R. A., Madore, B. F., van den Bosch, F. C. \& Balcells, M. 2003, \apj, {\bf 583}, 732

Tonini, C., Lapi, A. \& Salucci, P. 2006, \apj, {\bf 649}, 591

Tremaine, S. \& Weinberg, M. D. 1984, \mnras, {\bf 209}, 729

Valenzuela, O. \& Klypin, A. 2003, \mnras, {\bf 345}, 406

Valenzuela, O., Rhee, G., Klypin, A., Governato, F., Stinson, G., Quinn, T. \& Wadsley, J. 2007, \apj, {\bf 657}, 773

Wechsler, R. H., Bullock, J. S., Primack, J. R., Kravtsov, A. V. \& Dekel, A. 2002, \apj, {\bf 568}, 52

Weinberg, M. D. \& Katz, N. 2002, \apj, {\bf 580}, 627

Weinberg, M. D. \& Katz, N. 2007a, \mnras, {\bf 375}, 425

Weinberg, M. D. \& Katz, N. 2007b, \mnras, {\bf 375}, 460

Weiner, B. J., Sellwood, J. A. \& Williams, T. B. 2001, \apj, {\bf 546}, 931

Weiner, B. J. 2004, in IAU Symp.\ 220, Dark Matter in Galaxies, ed.\ S. Ryder, D. J. Pisano, M. Walker \& K. C. Freeman (Dordrecht: Reidel), p.~35

Z\'anmar S\'anchez, R., Sellwood, J. A., Weiner B. J. \& Williams, T. B. 2008, \apj, {\bf 674}, 797

\end{document}